\documentclass[12pt,a4paper,titlepage]{article}
\usepackage[utf8]{inputenc}
\usepackage[top=50pt,bottom=50pt,left=68pt,right=66pt]{geometry}
\usepackage{amsmath}
\usepackage{booktabs}
\usepackage{tikz}
\usetikzlibrary{arrows, calc, positioning}
\usetikzlibrary{arrows.meta}
\usepackage{amssymb}
\usepackage[title]{appendix}
\usepackage{mathrsfs}
\usepackage{float}
\usepackage{multirow}
\usepackage{graphicx,caption,subcaption}
\usepackage[space]{grffile}
\usepackage{cite}

\begin{document}
\renewcommand{\arraystretch}{1.3}

\makeatletter
\def\@hangfrom#1{\setbox\@tempboxa\hbox{{#1}}%
      \hangindent 0pt
      \noindent\box\@tempboxa}
\makeatother


\def\un#1{\relax\ifmmode\@@underline#1\else
        $\@@underline{\hbox{#1}}$\relax\fi}


\let\under=\unt                 
\let\ced=\ce                    
\let\du=\du                     
\let\um=\Hu                     
\let\sll=\lp                    
\let\Sll=\Lp                    
\let\slo=\os                    
\let\Slo=\Os                    
\let\tie=\ta                    
\let\br=\ub                     


\def\a{\alpha}
\def\b{\beta}
\def\c{\chi}
\def\d{\delta}
\def\e{\epsilon}
\def\f{\phi}
\def\g{\gamma}
\def\h{\eta}
\def\i{\iota}
\def\j{\psi}
\def\k{\kappa}
\def\l{\lambda}
\def\m{\mu}
\def\n{\nu}
\def\o{\omega}
\def\p{\pi}
\def\q{\theta}
\def\r{\rho}
\def\s{\sigma}
\def\t{\tau}
\def\u{\upsilon}
\def\x{\xi}
\def\z{\zeta}
\def\D{\Delta}
\def\F{\Phi}
\def\G{\Gamma}
\def\J{\Psi}
\def\L{\Lambda}
\def\O{\Omega}
\def\P{\Pi}
\def\Q{\Theta}
\def\S{\Sigma}
\def\U{\Upsilon}
\def\X{\Xi}


\def\ve{\varepsilon}
\def\vf{\varphi}
\def\vr{\varrho}
\def\vs{\varsigma}
\def\vq{\vartheta}


\def\ca{{\cal A}}
\def\cb{{\cal B}}
\def\cc{{\cal C}}
\def\cd{{\cal D}}
\def\ce{{\cal E}}
\def\cf{{\cal F}}
\def\cg{{\cal G}}
\def\ch{{\cal H}}
\def\ci{{\cal I}}
\def\cj{{\cal J}}
\def\ck{{\cal K}}
\def\cl{{\cal L}}
\def\cm{{\cal M}}
\def\cn{{\cal N}}
\def\co{{\cal O}}
\def\cp{{\cal P}}
\def\cq{{\cal Q}}
\def\car{{\cal R}}
\def\cs{{\cal S}}
\def\ct{{\cal T}}
\def\cu{{\cal U}}
\def\cv{{\cal V}}
\def\cw{{\cal W}}
\def\cx{{\cal X}}
\def\cy{{\cal Y}}
\def\cz{{\cal Z}}


\def\Sc#1{{\hbox{\sc #1}}}      
\def\Sf#1{{\hbox{\sf #1}}}      



\def\slpa{\slash{\pa}}                            
\def\slin{\SLLash{\in}}                                   
\def\bo{{\raise-.3ex\hbox{\large$\Box$}}}               
\def\cbo{\Sc [}                                         
\def\pa{\partial}                                       
\def\de{\nabla}                                         
\def\dell{\bigtriangledown}                             
\def\su{\sum}                                           
\def\pr{\prod}                                          
\def\iff{\leftrightarrow}                               
\def\conj{{\hbox{\large *}}}                            
\def\ltap{\raisebox{-.4ex}{\rlap{$\sim$}} \raisebox{.4ex}{$<$}}   
\def\gtap{\raisebox{-.4ex}{\rlap{$\sim$}} \raisebox{.4ex}{$>$}}   
\def\TH{{\raise.2ex\hbox{$\displaystyle \bigodot$}\mskip-4.7mu \llap H \;}}
\def\face{{\raise.2ex\hbox{$\displaystyle \bigodot$}\mskip-2.2mu \llap {$\ddot
        \smile$}}}                                      
\def\dg{\sp\dagger}                                     
\def\ddg{\sp\ddagger}                                   

\font\tenex=cmex10 scaled 1200


\def\sp#1{{}^{#1}}                              
\def\sb#1{{}_{#1}}                              
\def\oldsl#1{\rlap/#1}                          
\def\slash#1{\rlap{\hbox{$\mskip 1 mu /$}}#1}      
\def\Slash#1{\rlap{\hbox{$\mskip 3 mu /$}}#1}      
\def\SLash#1{\rlap{\hbox{$\mskip 4.5 mu /$}}#1}    
\def\SLLash#1{\rlap{\hbox{$\mskip 6 mu /$}}#1}      
\def\PMMM#1{\rlap{\hbox{$\mskip 2 mu | $}}#1}   %
\def\PMM#1{\rlap{\hbox{$\mskip 4 mu ~ \mid $}}#1}       %
\def\Tilde#1{\widetilde{#1}}                    
\def\Hat#1{\widehat{#1}}                        
\def\Bar#1{\overline{#1}}                       
\def\sbar#1{\stackrel{*}{\Bar{#1}}}             
\def\bra#1{\left\langle #1\right|}              
\def\ket#1{\left| #1\right\rangle}              
\def\VEV#1{\left\langle #1\right\rangle}        
\def\abs#1{\left| #1\right|}                    
\def\leftrightarrowfill{$\mathsurround=0pt \mathord\leftarrow \mkern-6mu
        \cleaders\hbox{$\mkern-2mu \mathord- \mkern-2mu$}\hfill
        \mkern-6mu \mathord\rightarrow$}
\def\dvec#1{\vbox{\ialign{##\crcr
        \leftrightarrowfill\crcr\noalign{\kern-1pt\nointerlineskip}
        $\hfil\displaystyle{#1}\hfil$\crcr}}}           
\def\dt#1{{\buildrel {\hbox{\LARGE .}} \over {#1}}}     
\def\dtt#1{{\buildrel \bullet \over {#1}}}              
\def\der#1{{\pa \over \pa {#1}}}                
\def\fder#1{{\d \over \d {#1}}}                 


\def\fracmm#1#2{{\textstyle{#1\over\vphantom2\smash{\raise.20ex
        \hbox{$\scriptstyle{#2}$}}}}}                   
\def\half{\frac12}                                        
\def\sfrac#1#2{{\vphantom1\smash{\lower.5ex\hbox{\small$#1$}}\over
        \vphantom1\smash{\raise.4ex\hbox{\small$#2$}}}} 
\def\bfrac#1#2{{\vphantom1\smash{\lower.5ex\hbox{$#1$}}\over
        \vphantom1\smash{\raise.3ex\hbox{$#2$}}}}       
\def\afrac#1#2{{\vphantom1\smash{\lower.5ex\hbox{$#1$}}\over#2}}    
\def\partder#1#2{{\partial #1\over\partial #2}}   
\def\parvar#1#2{{\d #1\over \d #2}}               
\def\secder#1#2#3{{\partial^2 #1\over\partial #2 \partial #3}}  
\def\on#1#2{\mathop{\null#2}\limits^{#1}}               
\def\bvec#1{\on\leftarrow{#1}}                  
\def\oover#1{\on\circ{#1}}                              

\def\[{\lfloor{\hskip 0.35pt}\!\!\!\lceil}
\def\]{\rfloor{\hskip 0.35pt}\!\!\!\rceil}
\def\Lag{{\cal L}}
\def\du#1#2{_{#1}{}^{#2}}
\def\ud#1#2{^{#1}{}_{#2}}
\def\dud#1#2#3{_{#1}{}^{#2}{}_{#3}}
\def\udu#1#2#3{^{#1}{}_{#2}{}^{#3}}
\def\calD{{\cal D}}
\def\calM{{\cal M}}

\def\szet{{${\scriptstyle \b}$}}
\def\ulA{{\un A}}
\def\ulM{{\underline M}}
\def\cdm{{\Sc D}_{--}}
\def\cdp{{\Sc D}_{++}}
\def\vTheta{\check\Theta}
\def\fracm#1#2{\hbox{\large{${\frac{{#1}}{{#2}}}$}}}
\def\ha{{\fracmm12}}
\def\tr{{\rm tr}}
\def\Tr{{\rm Tr}}
\def\itrema{$\ddot{\scriptstyle 1}$}
\def\ula{{\underline a}} \def\ulb{{\underline b}} \def\ulc{{\underline c}}
\def\uld{{\underline d}} \def\ule{{\underline e}} \def\ulf{{\underline f}}
\def\ulg{{\underline g}}
\def\items#1{\\ \item{[#1]}}
\def\ul{\underline}
\def\un{\underline}
\def\fracmm#1#2{{{#1}\over{#2}}}
\def\footnotew#1{\footnote{\hsize=6.5in {#1}}}
\def\low#1{{\raise -3pt\hbox{${\hskip 0.75pt}\!_{#1}$}}}

\def\Dot#1{\buildrel{_{_{\hskip 0.01in}\bullet}}\over{#1}}
\def\dt#1{\Dot{#1}}

\def\DDot#1{\buildrel{_{_{\hskip 0.01in}\bullet\bullet}}\over{#1}}
\def\ddt#1{\DDot{#1}}

\def\DDDot#1{\buildrel{_{_{\hskip 0.01in}\bullet\bullet\bullet}}\over{#1}}
\def\dddt#1{\DDDot{#1}}

\def\DDDDot#1{\buildrel{_{_{\hskip 
0.01in}\bullet\bullet\bullet\bullet}}\over{#1}}
\def\ddddt#1{\DDDDot{#1}}

\def\Tilde#1{{\widetilde{#1}}\hskip 0.015in}
\def\Hat#1{\widehat{#1}}


\newskip\humongous \humongous=0pt plus 1000pt minus 1000pt
\def\caja{\mathsurround=0pt}
\def\eqalign#1{\,\vcenter{\openup2\jot \caja
        \ialign{\strut \hfil$\displaystyle{##}$&$
        \displaystyle{{}##}$\hfil\crcr#1\crcr}}\,}
\newif\ifdtup
\def\panorama{\global\dtuptrue \openup2\jot \caja
        \everycr{\noalign{\ifdtup \global\dtupfalse
        \vskip-\lineskiplimit \vskip\normallineskiplimit
        \else \penalty\interdisplaylinepenalty \fi}}}
\def\li#1{\panorama \tabskip=\humongous                         
        \halign to\displaywidth{\hfil$\displaystyle{##}$
        \tabskip=0pt&$\displaystyle{{}##}$\hfil
        \tabskip=\humongous&\llap{$##$}\tabskip=0pt
        \crcr#1\crcr}}
\def\eqalignnotwo#1{\panorama \tabskip=\humongous
        \halign to\displaywidth{\hfil$\displaystyle{##}$
        \tabskip=0pt&$\displaystyle{{}##}$
        \tabskip=0pt&$\displaystyle{{}##}$\hfil
        \tabskip=\humongous&\llap{$##$}\tabskip=0pt
        \crcr#1\crcr}}


\def\eV{\,{\rm eV}}
\def\keV{\,{\rm keV}}
\def\MeV{\,{\rm MeV}}
\def\GeV{\,{\rm GeV}}
\def\TeV{\,{\rm TeV}}
\def\sv{\left<\sigma v\right>}
\def\({\left(}
\def\){\right)}
\def\cm{{\,\rm cm}}
\def\K{{\,\rm K}}
\def\kpc{{\,\rm kpc}}
\def\beq{\begin{equation}}
\def\eeq{\end{equation}}
\def\bea{\begin{eqnarray}}
\def\eea{\end{eqnarray}}


\newcommand{\be}{\begin{equation}}
\newcommand{\ee}{\end{equation}}
\newcommand{\nbe}{\begin{equation*}}
\newcommand{\nee}{\end{equation*}}

\newcommand{\fr}{\frac}
\newcommand{\lb}{\label}

\thispagestyle{empty}

{\hbox to\hsize{
\vbox{\noindent February 2025 \hfill IPMU25-0003 \\
\noindent  \hfill }}

\noindent
\vskip2.0cm
\begin{center}

{\large\bf One-loop corrections to the E-type $\alpha$-attractor models
of inflation and primordial black hole production}

\vglue.3in

Daniel Frolovsky~${}^{a,*}$ and Sergei V. Ketov~${}^{a,b,c,\#}$ 
\vglue.3in

${}^a$~Department of Physics and Interdisciplinary Research Laboratory, \\
Tomsk State University, Tomsk 634050, Russia\\
${}^b$~Department of Physics, Tokyo Metropolitan University, Tokyo 192-0397, Japan \\
${}^c$~Kavli Institute for the Physics and Mathematics of the Universe (WPI),
\\The University of Tokyo Institutes for Advanced Study,  Chiba 277-8583, Japan\\
\vglue.2in

${}^{*}$~frolovsky@mail.tsu.ru, ${}^{\#}$~ketov@tmu.ac.jp
\end{center}

\vglue.3in

\begin{center}
{\Large\bf Abstract}

\end{center}
\vglue.2in
\noindent The one-loop corrections (1LC) to the power spectrum of scalar perturbations arising from cubic interactions in the single-field E-type $\alpha$-attractor models of inflation and primordial black hole (PBH) production are numerically calculated. The results demonstrate the 1LC contributes merely a few percent to the tree-level power spectrum. The model parameters are chosen to predict the PBH masses in the asteroid-mass range, while maintaining consistency with the cosmic microwave background (CMB) observations within 1$\sigma$ confidence levels, and obeying the upper limits on $\mu$-distortions. The PBHs formed on scales smaller than the inflation scale can constitute a significant fraction of the present dark matter (DM). The PBH-induced gravitational waves (GW) may be detectable by the future space-based gravitational interferometers. We also consider a reconstruction of the scalar potential from possible GW observations and present a numerical approach tested in the model parameter space.

\newpage

\section{Introduction}

Cosmic microwave background (CMB) measurements \cite{Planck:2018jri} provide valuable insights into the early universe but do not uniquely fix the underlying model of inflation. Though the simplest single-field slow-roll (SR) models of inflation are tightly constrained by the 
CMB measurements and the Swampland Conjectures \cite{vanBeest:2021lhn} about their Ultra-Violet (UV) completion in quantum gravity,   significant uncertainties remain. Those uncertainties leave room for constructing more general models that not only describe inflation but also formation of primordial black holes (PBHs) from gravitational collapse of large scalar perturbations \cite{Novikov:1967tw,Hawking:1971ei}. 

The standard mechanism of PBH production in single-field models of inflation is via engineering a near-inflection point in an inflaton potential on energy scales under the scale of inflation, which cannot be probed by CMB  \cite{Karam:2022nym}. Inflaton dynamics is driven by the inflaton potential, whereas adding PBH production generically leads to lower values  of the CMB scalar tilt $n_s$, often in tension with CMB measurements. To get viable inflation with PBH production, and avoid tension with the CMB observations, fine-tuning of the inflaton potential parameters is needed. However, then another problem arises because the classical fine-tuning may be destroyed by quantum loop corrections, as was first observed in Ref.~\cite{Kristiano:2022maq}. The enhancement of the power spectrum of scalar perturbations, related to large perturbations and required for efficient PBHs production beyond the Hawking evaporation limit, has to be 6 or 7 orders of the magnitude higher against the CMB spectrum, which may imply large quantum corrections. Demanding a suppression of the loop corrections against the tree-level contribution can be used for an even tighter discrimination between the models of inflation with related PBHs production. 

The first version of Ref.~\cite{Kristiano:2022maq}, which attracted considerable attention in the literature 
\cite{Riotto:2023hoz,Choudhury:2023vuj,Kristiano:2023scm,Riotto:2023gpm,Firouzjahi:2023aum,Motohashi:2023syh,Firouzjahi:2023ahg,Fumagalli:2023hpa,Saburov:2024und,Inomata:2024lud,Braglia:2024zsl,Kristiano:2024vst,Kawaguchi:2024rsv,Fumagalli:2024jzz,Firouzjahi:2024sce}, claimed that {\it any} single-field model of inflation with PBH production suffers from a large one-loop correction (1LC) that destroys the viability of those models against the CMB observations, at least in the case of sharp transitions from slow-roll (SR) to ultra-slow-roll (USR) regimes. On the one hand, physical intuition tells us that physics on higher (energy) scales should be unaffected by lower (energy) scale physics, and effective zero modes should be unphysical for local observers inside the horizon, which implies small 1LC on the super-Hubble scales. On the other hand, the Wilsonian approach to effective field theory may be violated with gravity because infrared and ultraviolet degrees of freedom cannot be separated. Therefore, it is reasonable to calculate 1LC in particular models with specific inflaton potentials and different methods.

An important class of viable single-field models of inflation is given by the $\alpha$-attractors \cite{Kallosh:2013hoa, Galante:2014ifa}.
Those models are extensions of the Starobinsky model of inflation  \cite{Starobinsky:1980te}, which is recovered at $\a=1$, though with  a more general ($\a$-dependent) prediction for the CMB tensor-to-scalar ratio $r$. The basic $\alpha$-attractor models are divided into two types depending upon the global shape of the inflaton scalar potential,
\be \lb{EaT}
\text{E-type:}~V\sim \left(1-e^{-\sqrt{\frac{2}{3\a}}\,\f/M_{\rm Pl}}\right)^2~, \quad {\rm and} \quad \text{T-type:}~V\sim \tanh^2\frac{\f/M_{\rm Pl}}{\sqrt{6\a}}~,
\ee
as a function of canonical inflaton $\f$. Both the E-type and T-type $\a$-attractor models can be generalized further, without changing their predictions for
inflation, by using proper functions $f(y)$ and $g(\tilde{r})$, respectively,  with the potentials
\be \lb{gEaT}
\text{E-type:}~V\sim f^2(y)~, \quad {\rm and} \quad \text{T-type:}~V\sim g^2(\tilde{r})~,
\ee
where the variables  $y$ and $\tilde r$ are defined as 
\begin{equation} \lb{defy}
	y = \exp \left( -\sqrt{\frac{2}{3\alpha}} \f/M_{\rm Pl}\right)\,, \quad  \tilde r= \tanh \left(\frac{\f/M_{\rm Pl}}{\sqrt{6 \alpha}}\right)~.
\end{equation}
It is worth mentioning that the full-square-form of the potentials in Eq.~(\ref{gEaT}) allows their embedding into supergravity \cite{Ketov:2019toi}.

After expanding the functions $f(y)$ and $g(\tilde{r})$ in power series and keeping a few leading terms, one may engineer a large peak in the power spectrum of scalar perturbations, leading to PBHs production. In the case of the generalized T-models, this construction was first used in Ref.~\cite{Dalianis:2018frf}, where a quintessence model was proposed with the inflaton potential
\begin{equation}\label{tpot1}
V=V_0 \left[c_0+c_1 \tilde r +c_2 \tilde r^2 +c_3 \tilde r^3\right]^2~.
\end{equation}
Here, the constant $V_0>0$ is related to the amplitude of scalar perturbations, and the parameters $c_0, c_1, c_2,$ and $c_3$ are all dimensionless. Fine-tuning of the parameters allows PBHs production~\cite{Dalianis:2018frf}, though at the price of the low tilt $n_s$, while more precise (later) CMB measurements ruled out the model (\ref{tpot1}) before examining loop corrections.

Therefore, it makes sense to turn to the generalized E-models with PBHs production. To get a good (within 1 sigma) agreement  with the latest CMB measurements, it was proposed in Refs.~\cite{Frolovsky:2022qpg,Frolovsky:2023hqd} to consider the E-model with the potential
 \begin{equation} \label{epot1}
V(\phi) = \frac{3}{4}M^2M^2_{\rm Pl}\left[ 1 - y - \theta y^{-2} + y^2(\beta - \gamma y)\right]^2\,,
\end{equation}
where $M$ is the inflaton mass, and  $\beta,\gamma,\theta$ are the additional dimensionless parameters. In particular, it was found 
\cite{Frolovsky:2022qpg} that demanding a regular function $f(y)$ at the origin $y=0$ does not allow agreement with the Planck-measured
value of $n_s$ within 1 sigma in the cases, where  $f(y)$  facilitates PBH production across a wide  mass range. However, adding a negative power of  $y$  resolves this tension \cite{Frolovsky:2023hqd}. In contrast to the T-models (\ref{tpot1}), where adding the negative powers of $\tilde{r}$ would imply a singularity at $\f=0$, the E-models (\ref{epot1}) have the infinite potential  at $y=0$ and $y=+\infty$ that correspond to $\f=+\infty$ and $\f=-\infty$, respectively.

To further investigate the validity of the generalized E-models for inflation and PBH production, one has to evaluate an impact of the loop corrections in the case of (\ref{epot1}) because the amplified fluctuations associated with PBH production may lead to the loop corrections exceeding the tree-level contribution calculated by using the linear perturbation equations in Refs.~\cite{Frolovsky:2022qpg,Frolovsky:2023hqd}. Should those loop corrections be too large, it would indicate a breakdown of perturbativity in the model, thus challenging reliability of its predictions for the CMB observables. This issue has been the subject of intensive research in the recent literature 
\cite{Kristiano:2022maq,Riotto:2023hoz,Kristiano:2023scm,Riotto:2023gpm,Choudhury:2023vuj,Firouzjahi:2023aum,Motohashi:2023syh,Firouzjahi:2023ahg,Saburov:2024und,Inomata:2024lud,Kristiano:2024vst,Firouzjahi:2024sce,Braglia:2024zsl,Kawaguchi:2024rsv,Fumagalli:2024jzz}. In this paper, we numerically compute the one-loop correction (1LC) to the scalar power spectrum in the generalized E-type 
$\alpha$-attractor model (\ref{epot1}) by using the 1LC formula obtained in Refs.~\cite{Franciolini:2023agm,Davies:2023hhn}.

The paper is organized as follows. In Sec.~2 we reparameterize the model (\ref{epot1})  in terms of the new parameters directly related to the shape of the inflaton potential and explore the potential for various values of the parameter $\a$. The perturbative dynamics of inflaton during SR and USR is numerically evaluated in Sec.~3. The one-loop (1LC) correction is computed in Sec.~4. The PBH-production-induced gravitational waves (GW) are derived in Sec.~5, in the second order with respect to perturbations. The inverse procedure from the given GW spectrum back to the power spectrum and a subsequent reconstruction of the inflaton potential  are outlined in Sec.~6, where this procedure is applied to our E-model. In Sec.~7 we summarize our main results. Sec.~8 is our conclusion.

\section{The Model}

The standard (quintessence) action of single-field inflation models reads
\begin{equation}
\mathcal{S}=\int \mathrm{d}^4 x \sqrt{-g}\left(\frac{1}{2} R-\frac{1}{2}\left(\partial_\mu \phi\right)^2-V(\phi)\right), \quad c=\hbar=M_{\text{Pl}}=1,
\end{equation}
where $\phi$ is a canonical inflaton field, and $V(\phi)$ represents its potential.\footnote{We use the mostly-plus signature of spacetime and the standard notation of General Relativity.}

The scalar potential of the E-type $\alpha$-attractor model (\ref{epot1}) can be rewritten in terms of the new dimensionless parameters
$\phi_i$ and $\sigma$ defined by 
\begin{equation}\label{newpar}
	\beta = \frac{\exp \left( \sqrt{\frac{2}{3\alpha}} \phi_{i}\right)}{1-\sigma^2}~,  \quad {\rm and} \quad 
\gamma = \frac{\exp \left(2\sqrt{\frac{2}{3\alpha}} \phi_{i}\right)}{3(1-\sigma^2)}~,
\end{equation}
that have the simple geometrical meaning: at $\sigma = 0$ the potential has an inflection point at $\phi = \phi_i$, and when $0 <\sigma \ll 1$, the potential has a local minimum $y_{\text{ext}}^{-}$ to the right of the inflection point and a local maximum $y_{\text{ext}}^{+}$ to the left of the inflection point. Both extrema are symmetrically located around the inflection point as
 \begin{equation}\label{extr2} 
 	y^{\pm}_{\rm ext} = y_{i}\left( 1\pm \sigma\right)~.
 \end{equation}
 
The parameter $\theta$ governs the slope of the plateau in the potential on CMB scales, allowing us to achieve precise tuning of the scalar tilt  $n_s$, and the tensor-to-scalar ratio $r$ \cite{Frolovsky:2023hqd}. The potential  (\ref{epot1})  realizes the double inflation scenario with an USR phase between two SR regimes of inflation, leading to a large enhancement of the power spectrum of scalar perturbations and, thus, the PBHs production. The profiles of the scalar potential for various values of the $\a$-parameter are given in Fig.~\ref{fig1}.

\begin{figure}[ht]
\centering
\includegraphics[width=0.7\linewidth]{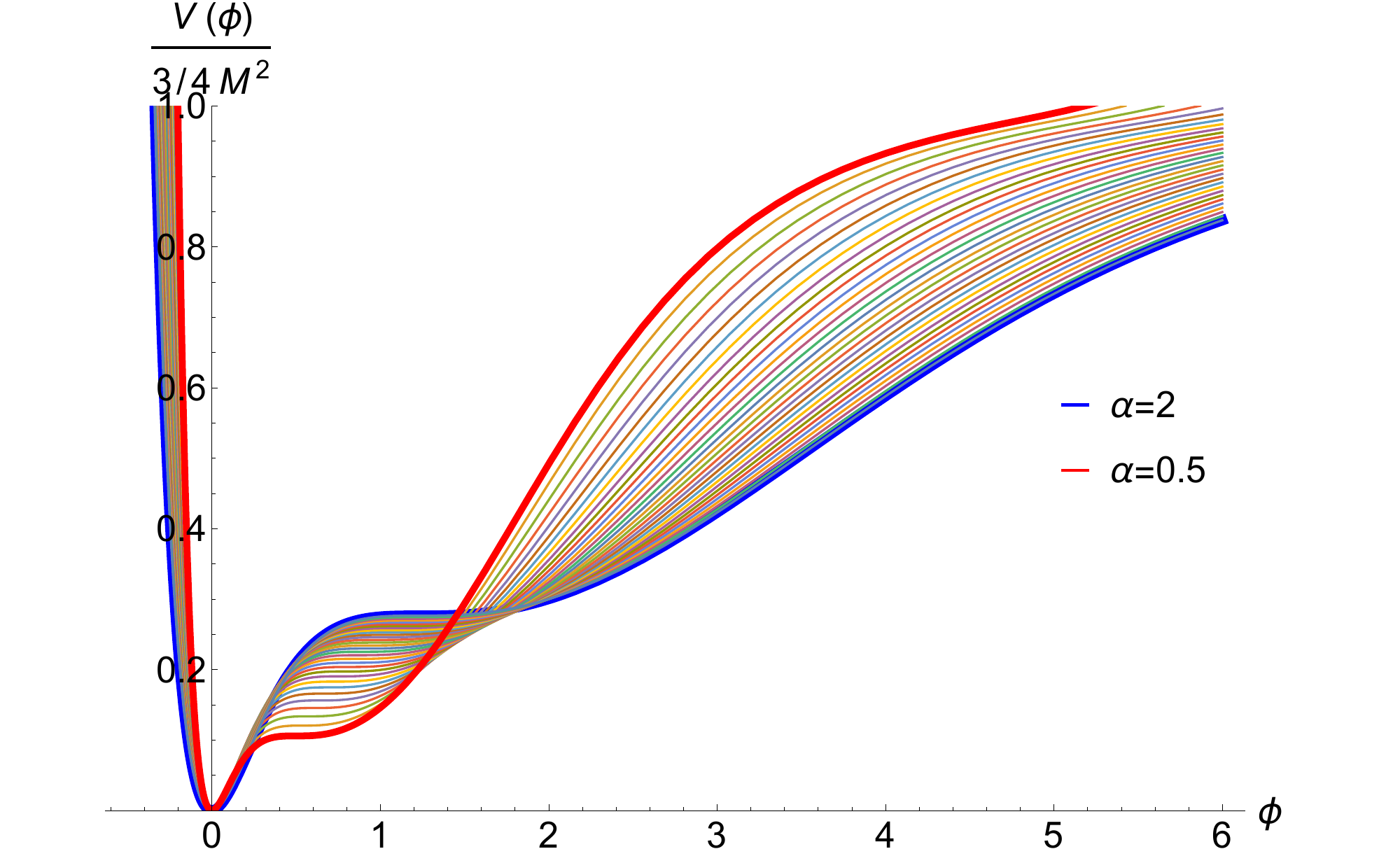}
\caption{The profiles of the scalar potential in the E-model for $\alpha \in [0.5, \ldots, 2]$. The scalar potential (\ref{epot1}) is shifted to get $V(0)=0$ for all $\alpha$. The other parameters of the model are given by set 1 of Table 1 in Section 7.}
\lb{fig1}
\end{figure}

\section{Background evolution  and curvature perturbations}

The evolution of inflaton field in a (flat) Friedmann-Lemaitre-Robertson-Walker (FLRW) universe is described by the standard (Klein-Gordon and Friedmann) equations: 
\begin{equation}\label{eomt}
\ddot{\phi}+3 H \dot{\phi}+V_{,\phi}=0~, \quad 3 H^2=\frac{1}{2} \dot{\phi}^2+V(\phi)~, \quad \dot{H}=-\frac{1}{2} \dot{\phi}^2~,
\end{equation}
where $H(t)\equiv \dot{a}/a$ is Hubble function, $a(t)$ is the cosmic scale factor in the (flat) FLRW metric, 
$ds^2=-dt^2 +a^2(dx_1^2+dx_2^2+dx_3^2)$, $V_{,\phi} =dV/d \phi$, and the dots denote the time derivatives. It is more convenient to use e-folds instead of time, which are defined by
\begin{equation}
	N=\int H(t)\,dt~.
\end{equation}
Then, the equations of motion read
\begin{equation}
	H^2\phi^{\prime \prime} +HH^{\prime}\phi^{\prime}+3H^2\phi^{\prime}+V_{,\phi}=0 \,, \quad H^2=\frac{2V(\phi)}{6-(\phi^{\prime})^2}~, \quad H^{\prime}=-\frac{1}{2}(\phi^{\prime})^2H~,
\end{equation}
or, equivalently,
\begin{equation}\label{eom}
	\phi^{\prime \prime}+3\phi^{\prime}-\frac{1}{2}(\phi^{\prime})^3+ \bigg(3-\frac{1}{2}(\phi^{\prime})^2\bigg)\frac{V_{,\phi}}{V}=0~,
\end{equation}
where the primes denote the derivatives with respect to e-folds $N$.

The standard Hubble-flow parameters are defined by
\begin{equation}
	\epsilon_{i+1}=\epsilon^{\prime}_{i}/\epsilon_i~, \quad  \quad \epsilon_0=H^{-1}.
\end{equation}

To distinguish the different phases of inflation and compute the one-loop quantum corrections (1LC), we use the first three Hubble-flow parameters,
\begin{equation}
\epsilon\equiv\epsilon_1~,\quad \eta\equiv\epsilon_2~, \quad \xi\equiv\epsilon_3~. 
\end{equation}
 The SR phase is characterized by the condition $\epsilon\ll|\eta|\ll1$, whereas the USR phase is defined by $|\eta|\approx6$. A transition between SR and USR phases is supposed to be continuous and not instantaneous. For the purpose of evaluating the 1LC, exact values of e-folds at the start and the end of the USR phase are needed. To determine the transition periods, we use the condition proposed in Ref.~\cite{Davies:2023hhn}:
  \begin{equation} \lb{bc}
 	|\xi|\geq1~.
 \end{equation}
 The results of our numerical calculation are given by Fig.~\ref{fig2}.
 
 \begin{figure}[ht]
\centering
\includegraphics[width=0.7\linewidth]{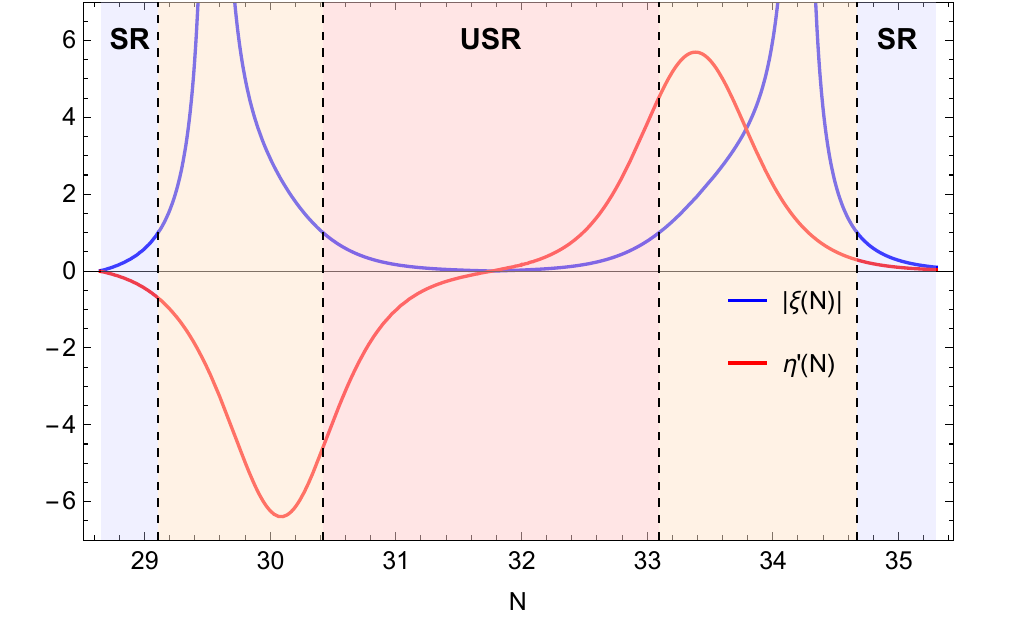}
\caption{The evolution of $\eta^{\prime}$ and $|\xi|$ as the functions of e-folds in the E-model. The vertical dashed lines mark the transitions where $|\xi|=1$. The parameters of the model are given by set 1 of Table 1 in Section 7.}
\lb{fig2}
\end{figure}

The start $N_s$  and the end  $N_e$ of the USR phase, obtained from the condition $|\xi|=1$, are directly associated with  the SR--USR and USR--SR transitions, respectively.  One can also introduce the (transition) sharpness parameter $h$ according to Ref.~\cite{Cai:2018dkf} as
  \begin{equation}
  	h\equiv6\frac{\sqrt{\epsilon_V}}{\pi_e}~,
  \end{equation}
where $\epsilon_V=V_{,\phi}(\phi_e)/(2V(\phi_e))$ is the standard slow-roll parameter, $\phi_e=\phi(N_e)$, and $\pi_e=\phi^{\prime}(N_e)$. In the case of a sharp transition, $h\leq-\,6$, and for a smooth transition $h\rightarrow 0$.

The standard expression for the power spectrum of scalar perturbations in the SR  approximation is given by
\begin{equation}\label{PSsr}
\mathcal{P}_R = \frac{H^2}{8 \pi^2 \epsilon}~,
\end{equation}
that is the simple tool for analyzing a dependence of the power spectrum on the model parameters. The shape of the power spectrum for 
some values of the parameter $\alpha$, obtained by numerical calculations based on (\ref{PSsr}), is given by Fig.~\ref{fig3}.

\begin{figure}[ht]
\centering
\includegraphics[width=0.7\linewidth]{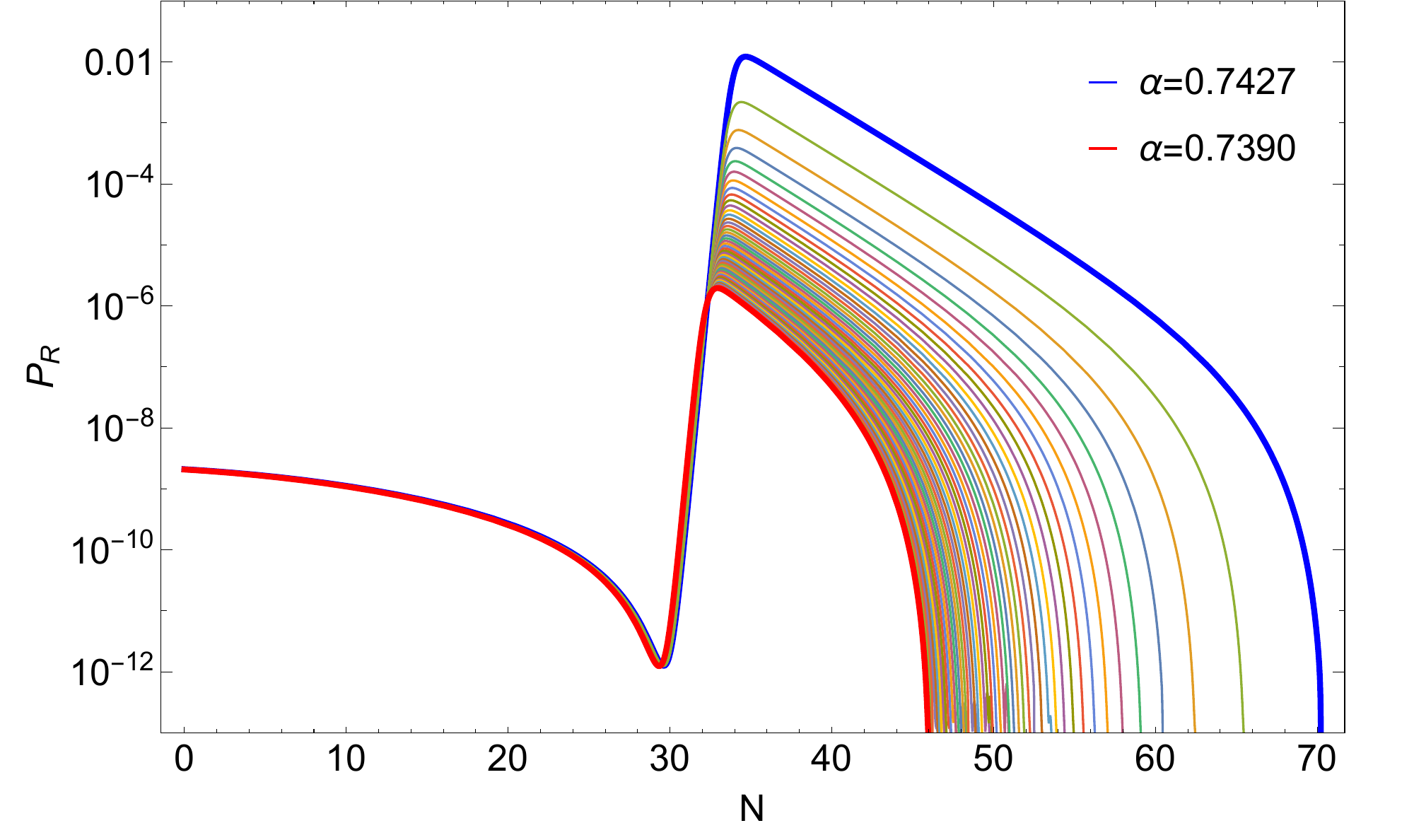}
\caption{The profiles of the power spectrum of scalar perturbations in the SR approximation in the E-model for $\alpha \in [0.7427, \ldots, 0.7390]$ with small and evenly spaced steps. For each value of $\alpha$, the scalar potential is shifted by an additive constant to get $V(0)=0$. The other parameters are given by set 1 of Table 1 in Section 7.}
\lb{fig3}
\end{figure}

As is evident from Fig.~\ref{fig3}, the shape of the power spectrum is sensitive to changes in $\alpha$. When $\alpha$ increases, the peak amplitude of the spectrum also grows. Furthermore, sensitivity of the peak amplitude against changes of $\alpha$ increases with larger values of $\alpha$. This relation is illustrated in Fig.~\ref{fig4}.
 
\begin{figure}[ht]
\centering
\includegraphics[width=0.7\linewidth]{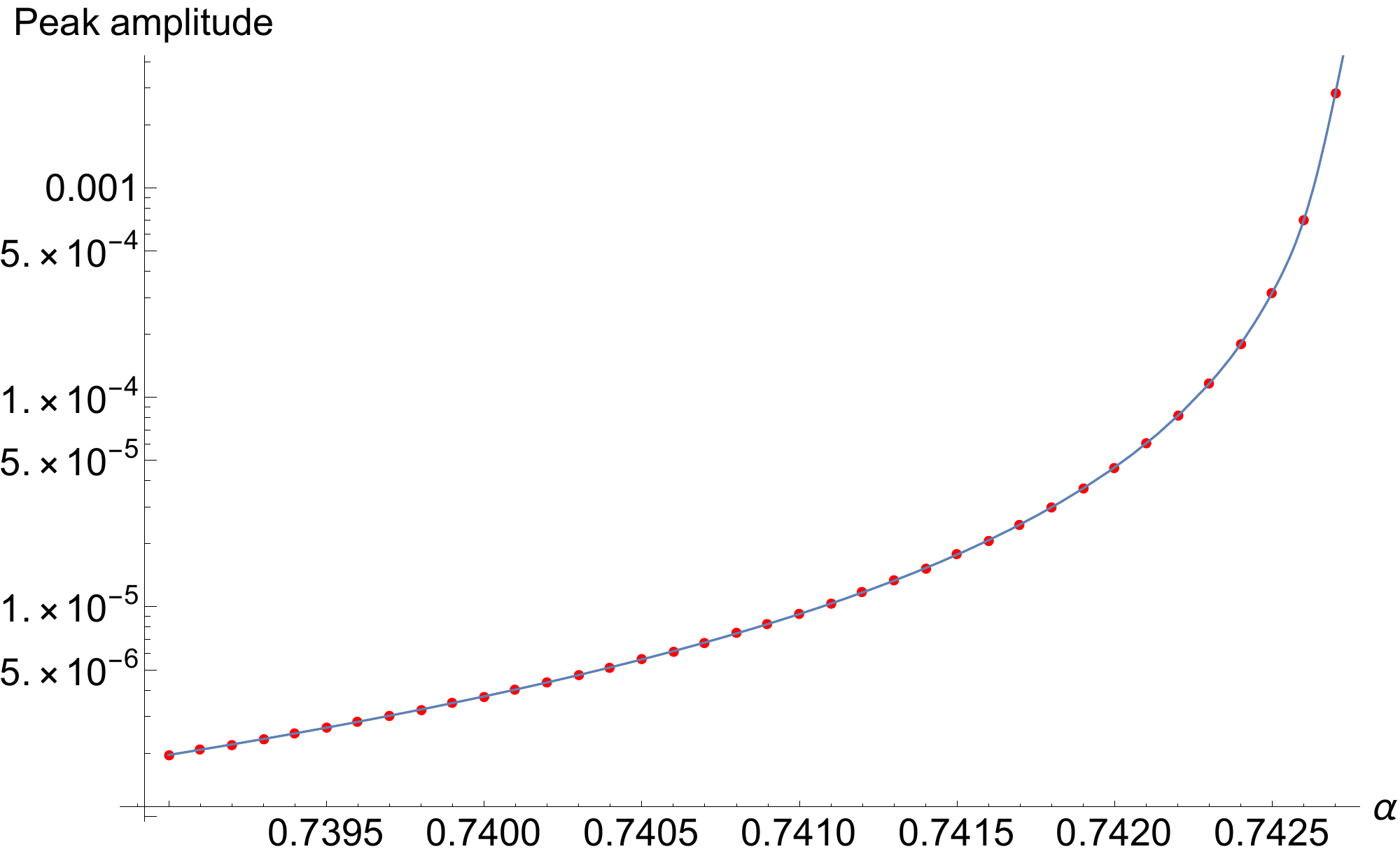}
\caption{The amplitude of the peak in the power spectrum of scalar perturbations in the SR approximation for the E-model in the logarithmic scale with $\alpha \in [0.7427, \ldots, 0.7390]$ in small and evenly spaced steps. The other parameters of the model are given by set 1 of Table 1 in Section 7.}
\label{fig4}
\end{figure}

A curvature perturbation during inflation, $\zeta_k(N)$,  can be expressed in terms of  $v_k$ as $\zeta_k\equiv v_k/z$, where $z\equiv a\cdot\phi^{\prime}$. The variable $v_k$ obeys the  Mukhanov-Sasaki (MS) equation
\begin{equation}\label{ms}
	v_k^{\prime \prime}+\left(1-\epsilon\right) v_k^{\prime}+\left[\frac{k^2}{a^2 H^2}+(1+\eta/2)(\epsilon- \eta/2 -2)-\eta^{\prime}/2 \right] v_k=0~.
\end{equation}
We solve this equation separately for its real and imaginary parts. For each mode, an integration starts five e-folds before the horizon crossing. In terms of the real and imaginary parts of $v_k$, the initial (Bunch-Davies) conditions are
 \begin{equation}\label{mseq}
 	\text{Re} (v_k)=\frac{1}{\sqrt{2 k}}~, \quad 	\text{Re}(v_k^{\prime})=0~, \quad \text{Im}(v_k)=0~, \quad \text{Im}(v_k^{\prime})=-\frac{i}{k_{\text{in}}}\sqrt{\frac{k}{2}}~,
 \end{equation}
 where $k_{in}$  is the mode that crossed the horizon at the start of integration.
 
Based on solutions to Eq.~(\ref{mseq}), the exact scalar power spectrum is computed as
\begin{equation}\label{ps}
	\mathcal{P}_\zeta(k)=\frac{k^3}{2 \pi^2}\Big|\frac{v_k}{z}\Big|^2~,
\end{equation}
where each mode is evaluated at the end of inflation.

The power spectrum of scalar perturbations can be approximated by the broken power-law fit as
\begin{equation} \lb{bpl}
	\mathcal{P}_{\text{BPL}}=A\frac{\alpha_1+\beta_1}{\beta_1(k/k_*)^{-\alpha_1}+\alpha_1(k/k_*)^{\beta_1}}~,
\end{equation}
where the parameters $\alpha_1$ and $\beta_1$ describe growth and decay of the spectrum around its peak, respectively. The profiles described by Eq.~(\ref{bpl}) often arise in the single-field inflation models with USR between two SR \cite{Byrnes:2018txb,Carrilho:2019oqg,Vaskonen:2020lbd,Tasinato:2020vdk,LISACosmologyWorkingGroup:2024hsc,Li:2024lxx,Cielo:2024poz}. As is illustrated by Fig.~\ref{fig5} obtained from our numerical calculations, the broken power law fit (\ref{bpl}) gives a good approximation for the peak of the power spectrum in the E-model.

\begin{figure}[ht]
\centering
\includegraphics[width=0.7\linewidth]{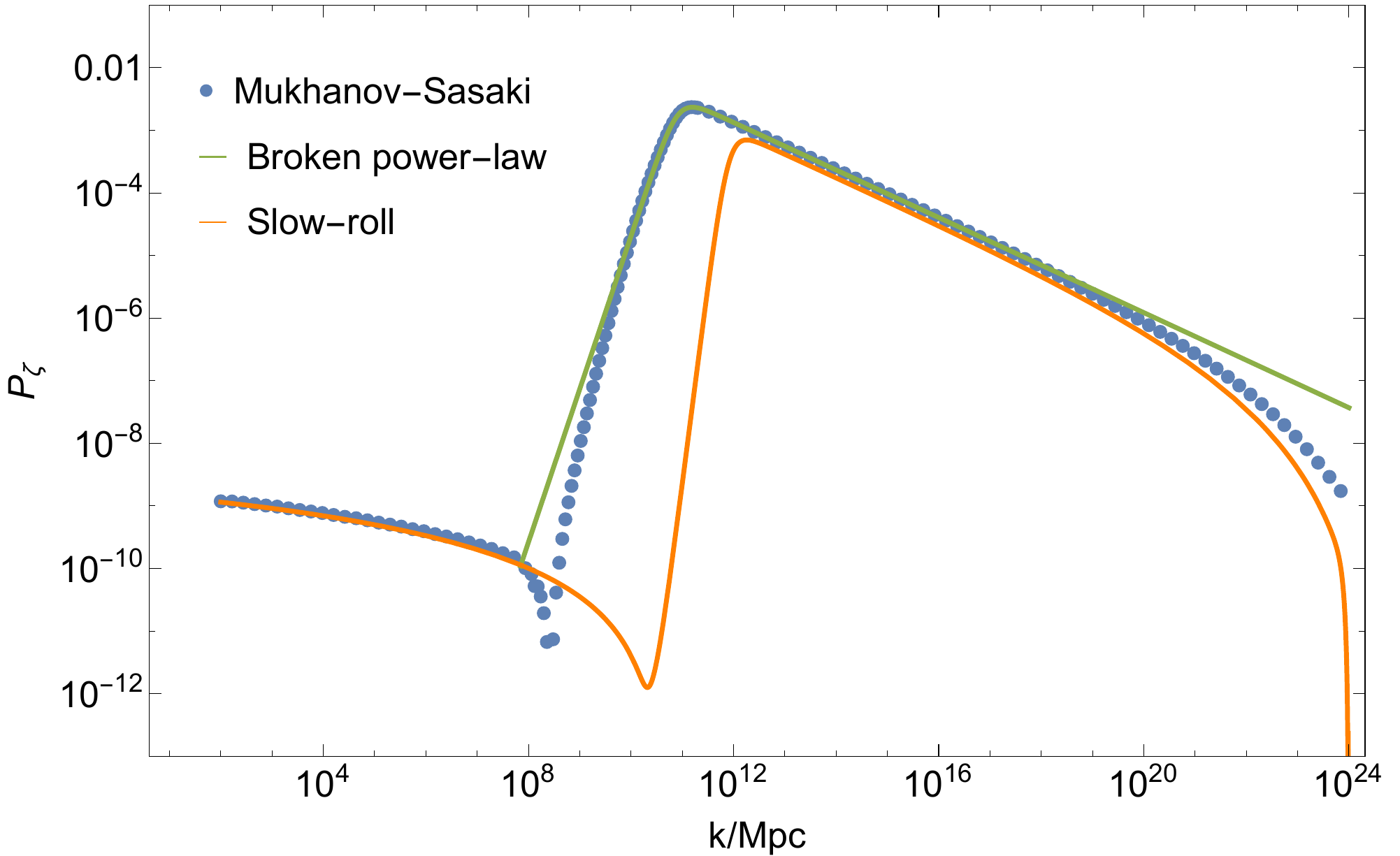}
\caption{The exact power spectrum of scalar perturbations derived from the MS equation (\ref{mseq}) together  with the broken power-law (BPL) fit and the slow-roll approximation in the E-model. The BPL fit parameters are $A = 0.00233$, $\alpha_1 = 2.446$, $\beta_1 = 0.379$, and $k_* = 1.57 \cdot 10^{11}$. The parameters of the E-model are given by set 1 of Table 1 in Section 7.}
\lb{fig5}
\end{figure}

\section{One-loop quantum correction in the E-model}

For a numerical calculation of the one-loop correction (1LC)  to the scalar power spectrum, we adopted the result from Refs.~\cite{Franciolini:2023agm,Davies:2023hhn} that takes into account cubic interactions and reads
\begin{equation}\label{1LC}
	\begin{aligned} \mathcal{P}_\zeta(p ; N)_{1\text {L}}=\frac{p^3}{\pi^4} \int\limits_{N_i}^{N_f} \mathrm{~d} N_1\, \epsilon(N_1) \frac{\mathrm{d} \eta(N_1)}{\mathrm{d}N_1}\big(a(N_1)\big)^2 \int\limits_{N_i}^{N_1} \mathrm{~d} N_2\, \epsilon(N_2) \frac{\mathrm{d}\eta(N_2)}{\mathrm{d} N_2}\big(a(N_2)\big)^2 \\ \int\limits_{k_s}^{k_e} \mathrm{~d} k\, k^2 \,\text{Im}\Bigg[\zeta_p(N)\,a(N_1)\,H(N_1)\,a(N_2)\,H(N_2)\,\frac{\mathrm{d}}{\mathrm{d} N_2}\left[\zeta_p^*(N_2)\big(\zeta_k^*(N_2)\big)^2\right] \\ \times\left\{\zeta_p^I(N) \frac{\mathrm{d}}{\mathrm{d} N_1}\left[\zeta_p^R(N_1) \big(\zeta_k(N_1)\big)^2\right]-\zeta_p^R(N)\,\frac{\mathrm{d}}{\mathrm{d} N_1}\left[\zeta_p^I(N_1) \big(\zeta_k(N_1)\big)^2\right]\right\}\Bigg]~,
	\end{aligned}
\end{equation}
where the large-scale mode $p$ is supposed to be located in the plateau region, being sufficiently separated from the peak scale $k_{\rm peak}$. We take $p=k_{\rm peak}/10^5$ in our E-model.  The e-folds $N$ are chosen well after the mode $p$ crosses the horizon $p\ll aH$, when $\zeta_p$ approaches a constant value. In (\ref{1LC}), $\zeta_p^R$ and $\zeta_p^I$ denote real and imaginary part of $\zeta_p$, respectively. The integration over the e-folds in (\ref{1LC}) is supposed to be performed over the period when $\eta^{\prime}$ is large, thus ensuring all relevant contributions to the 1LC are captured. Therefore, the integration bounds are defined from $N_i$, around an e-fold before the first transition, to $N_f$, around an e-fold after the last transition. The loop integration over $k$ is supposed to be limited to the USR phase with $k_s=a(N_s)H(N_s)$ and $k_e=a(N_e)H(N_e)$. 

In particular, a derivation of Eq.~(\ref{1LC}) in Ref.~\cite{Davies:2023hhn} used the same method as in Ref.~\cite{Kristiano:2022maq} but evaluated the loop numerically without assuming instantaneous transitions. Equation  (\ref{1LC})  is suitable to our purposes because it allows us to compute the 1LC starting from the scalar potential. We used an original code for evaluating Eq.~(\ref{1LC}) in the E-model.

 We define the relative 1LC against the tree-level scalar power spectrum as
 \begin{equation}
 	\delta_{\rm 1L}=\frac{\mathcal{P}_\zeta(p ; N)_{1\text{L}}}{\mathcal{P}_\zeta(p)}\cdot 100\,\%~.
 \end{equation}

It follows from the results of our numerical calculations collected in Table 2 of Sec.~7  that the E-model has mild SR-USR transitions characterized by $|h|<6$. The 1LC from cubic and quartic interactions was also studied in Ref.~\cite{Firouzjahi:2023aum}, where it was shown that the cubic interaction dominates over the quartic one when $|h|<6$. In particular, according to Eqs.~(5.28) and (5.29) in Ref.~\cite{Firouzjahi:2023aum}, one has
 \begin{equation}
 \frac{\delta_{\rm H_3}}{\delta_{\rm H_4}}=\frac{108 h (h+8) \text{$\Delta $N}_{\text{USR}}}{6 (h (h+6)+36) \text{$\Delta $N}_{\text{USR}}-h+6}~.
 \end{equation}
 However, the value of $\delta_{\rm H_3}$ given above differs from our numerical results obtained from Eq.~(\ref{1LC}).  We believe this disagreement is due to the difference between sharp and mild transitions where large one-loop corrections are diminished during the subsequent evolution of  curvature perturbations after the USR phase.

In the E-model with the parameter set 1 in Table 1 and Table 2, we find  $\delta_{\rm H_3}/\delta_{\rm H_4}=-5.9$, so the cubic interaction is dominant indeed.

\section{PBH-production-induced gravitational waves}

Large scalar perturbations, whose amplitudes exceed the CMB amplitude $A_s$ by six or seven orders of magnitude (see Fig.~\ref{fig5}), may lead to a detectable stochastic GW background because tensor modes are sourced by scalar modes already in the second order with respect to perturbations, see Ref.~\cite{Domenech:2021ztg} for a review.

The present GW energy density computed in the second order is given by \cite{Kohri:2018awv,Espinosa:2018eve,Bartolo:2018evs},
\begin{equation}\label{gw}
\frac{\Omega_{\mathrm{GW}}(k)}{\Omega_r}=\frac{c_g}{72} \int\limits_{-\frac{1}{\sqrt{3}}}^{\frac{1}{\sqrt{3}}} \mathrm{d}\,d \int\limits_{\frac{1}{\sqrt{3}}}^{\infty} \mathrm{d}\,s\left[\frac{\left(s^2-\frac{1}{3}\right)\left(d^2-\frac{1}{3}\right)}{s^2+d^2}\right]^2 \mathcal{P}_\zeta(k x)\,\mathcal{P}_\zeta(k y)\,(I_c^2+I_s^2)~,
\end{equation}
where $\Omega_{r}$ is the present-day radiation density, $h^2\Omega_{r}\approx 4.2\cdot10^{-5}$  according to Ref.~\cite{Planck:2018jri}, normalized by the current Hubble parameter $h=0.674$ (if Hubble tension is ignored). The coefficient $c_g=0.4$ is related to
the number of the effective degrees of freedom in thermal radiation when assuming the  Standard Model physics.
The variables $x$ and $y$ are related to the integration variables as $x=\frac{\sqrt{3}}{2}(s+d)$, and $y=\frac{\sqrt{3}}{2}(s-d)$,
and the functions  $I_c$ and $I_s$ are given by \cite{Espinosa:2018eve,Bartolo:2018evs},
\begin{equation}
	\begin{aligned}  & I_s=-\,36 \frac{s^2+d^2-2}{\left(s^2-d^2\right)^2}\left[\frac{s^2+d^2-2}{s^2-d^2} \log \left|\frac{d^2-1}{s^2-1}\right|+2\right]~,\\ &  I_c=-\,36 \pi \frac{\left(s^2+d^2-2\right)^2}{\left(s^2-d^2\right)^3} \theta(s-1)~,\end{aligned}
\end{equation}
with the Heaviside step function $\theta$. It is worth noticing that the GW density profile obtained from Eq.(\ref{gw}) can be influenced by non-Gaussianity (see Eqs.~(16) and~(18) in Ref.~\cite{Iovino:2024sgs}), so  Eq.~(\ref{gw}) may acquire a prefactor, $\Omega_{\mathrm{GW}}(k) \rightarrow \mathcal{A}^4 \,\Omega_{\mathrm{GW}}(k)$. In our analysis, we neglect non-Gaussian effects.

Our results for the induced GW from a numerical calculation of Eq.~(\ref{gw}) in the E-model are given in Fig.~\ref{fig6} against the background of the expected sensitivity curves of the future space-based gravitational interferometers LISA \cite{LISA:2017pwj,LISACosmologyWorkingGroup:2024hsc,LISACosmologyWorkingGroup:2025vdz} and DECIGO \cite{Kudoh:2005as}.

\begin{figure}[ht]
\centering
\includegraphics[width=0.7\linewidth]{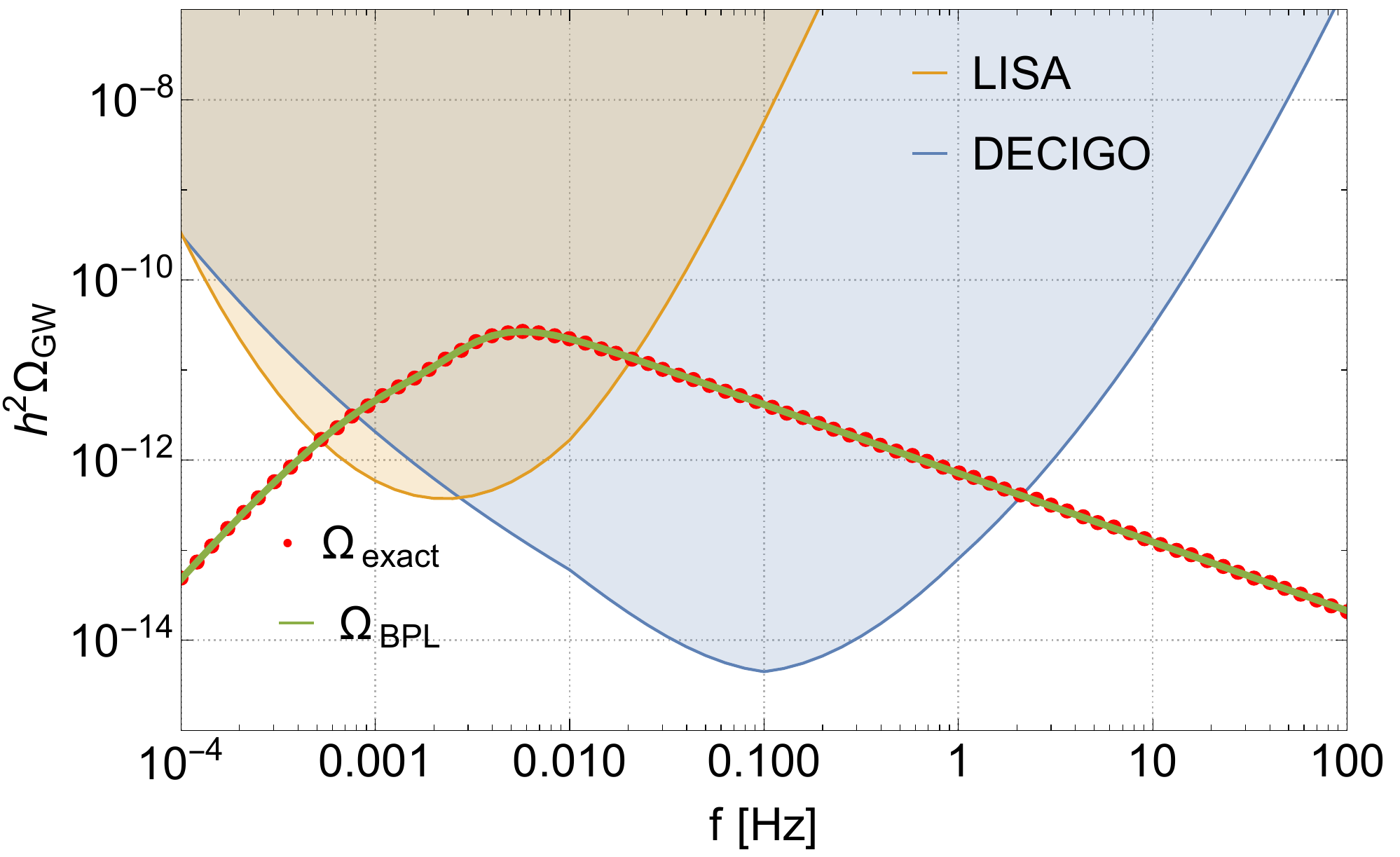}
\caption{The curve with red dots represents the GW density computed from Eq.~(\ref{gw}) by using the scalar perturbation spectrum 
$\mathcal{P}_\zeta(k)$ obtained from solving the MS equation (\ref{mseq}) with the model parameters of set 3 in Table 1 (see Sec.~7). The green line corresponds to the GW density derived by using the broken power-law approximation for the power spectrum. The orange and blue lines show the sensitivity curves of LISA and DECIGO, respectively. The frequency of GW is related to the comoving wavenumber $k$ as $k \simeq 6.47 \cdot 10^{14}\,({f}/\text{Hz})\,\mathrm{Mpc}^{-1}$.}
\lb{fig6}
\end{figure}

\section{Reconstruction of scalar potential from GW signal}

When assuming the source of a GW signal due to gravitational collapse of large scalar perturbations after inflation, it is natural to attempt  a reconstruction of the (single-field) inflation model underlying this signal. The reconstruction implies the following chain:
\begin{figure}[H]
\centering
\begin{tikzpicture}[node distance=2cm, auto]

\tikzstyle{box} = [
   rectangle,
   rounded corners,
   draw=black,
   minimum width=3cm,
   minimum height=1.2cm,
   text centered
]

\node[box] (gw) {
  \begin{tabular}{c}
    \textbf{GW signal} \\
    $\Omega_{\mathrm{GW}}(k)$ 
  \end{tabular}
};

\node[box, right=of gw] (ps) {
  \begin{tabular}{c}
    \textbf{Power spectrum} \\
    $\mathcal{P}_\zeta(k)$
  \end{tabular}
};

\node[box, right=of ps] (pot) {
  \begin{tabular}{c}
    \textbf{Scalar potential} \\
    $V(\phi)$
  \end{tabular}
};

\draw[->, line width=0.7pt, shorten >=3pt, shorten <=3pt, >=Stealth] (gw) -- (ps);
\draw[->, line width=0.7pt, shorten >=3pt, shorten <=3pt, >=Stealth] (ps) -- (pot);

\end{tikzpicture}
\label{fig:reconstruction}
\end{figure} 

This procedure is inevitably ambiguous, and the reconstruction  is still a longstanding challenge. For instance, the reconstruction formula proposed by Hodges and Blumenthal in Ref.~\cite{Hodges:1990bf},
\begin{equation}\lb{hb}
	\frac{1}{V(N)}=-\frac{1}{12 \pi^2 M_{\mathrm{Pl}}^4} \int \frac{d N}{\mathcal{P}_\zeta(N)}~,
\end{equation}
requires knowledge of the full power spectrum across various scales.

Different power spectra can yield almost identical GW energy density profiles over a limited range of scales, see, e.g., Fig. 1 and Fig. 2 in Ref. \cite{Frolovsky:2024zds}. Nevertheless, it is possible to systematically identify the power spectra corresponding to a specific GW energy density curve \cite{LISACosmologyWorkingGroup:2024hsc,LISACosmologyWorkingGroup:2025vdz}. Those spectra are often approximated by the broken power law or the  log-normal distribution. 

The observed CMB window to inflation is limited or small, revealing only the SR plateau at large scales. However, at smaller scales, future GW observations can shed light on the structure of the scalar potential by reconstructing the peak in the power spectrum. Once the peak is identified, the Hubble-flow parameters in its vicinity can be determined. For instance, the parameters $\alpha_1$ and $\beta_1$ in a broken power-law approximation are directly linked to the second slow-roll parameter $\eta$ \cite{Karam:2022nym}.  Using these parameters, the scalar potential at smaller scales can be reconstructed by using  the relations \cite{Byrnes:2018txb, Franciolini:2022pav}
\begin{equation}\label{recV}
	\begin{aligned}  & V(N)=V(N_{\text {ref }}) \exp \left\{-\,2 \int\limits_{N_{\text {ref }}}^N d N^{\prime}\left[\frac{\epsilon(3-\eta)}{3-\epsilon}\right]\right\}~, \\ & \phi(N)=\phi\left(N_{\text {ref }}\right) \pm \int\limits_{N_{\mathrm{ref}}}^N d N^{\prime} \sqrt{2 \epsilon}~.
	\end{aligned}
\end{equation}
However, the potential $V(\phi)$  still has to be determined as a function of  $\phi$  over all scales.

We propose to upgrade the reconstruction chain by incorporating theoretically well-motivated models of inflation. This approach focuses on a reconstruction of the parameters of the inflaton potential that are consistent with the reconstructed power spectrum. By doing so, it is possible to identify consistent theoretical models that can explain a given GW signal, while obeying phenomenological constraints across all scales. This process can be performed numerically.

To illustrate this approach, let us consider the reconstructed power spectrum  $\mathcal{P}^{\text{rec}}_\zeta(k)$ derived from a hypothetical GW signal. The goal is to determine the set of parameters $\vec{\theta}$  in a scalar potential $ V(\phi, \vec{\theta})$  that generates the power spectrum  $\mathcal{P}_\zeta(k, \vec{\theta})$ fitting the reconstructed one,  $\mathcal{P}_\zeta(k, \vec{\theta}) \approx \mathcal{P}_\zeta^{\text{rec}}(k)$. This task can be formalized as the minimization problem for the functional
\begin{equation}
	S(\vec{\theta}) = \sum_{i} \left[ \mathcal{P}_\zeta(k_i; \vec{\theta}) - \mathcal{P}^{\text{\rm rec}}_\zeta(k_i) \right]^2~,
\end{equation}
and then solved  by using the Levenberg-Marquardt algorithm. This method requires an initial guess for $\vec{\theta}$ that can be obtained numerically.

First, we find the function $\phi_{\text{sol}}[N, \vec{\theta}]$  as a numerical solution to the equation of motion~(\ref{eom}) with a given set $\vec{\theta}$. Based on this solution, the power spectrum can be calculated in the form (\ref{PSsr}). This allows us to manually adjust $\vec{\theta}$ matching the power spectrum in the SR approximation with $\mathcal{P}^{\text{rec}}_\zeta(k)$. Fine-tuning at this stage is not necessary, while it suffices to find $\vec{\theta}$ producing a peak in the power spectrum within the range of two or three orders of magnitude in $k$ around the peak of $\mathcal{P}^{\text{rec}}_\zeta(k)$. The initial guess for $\vec{\theta}$ can also be guided by analyzing the potential reconstructed from (\ref{recV}).
 
 After that, we numerically construct the function $v[k, \vec{\theta}]$ providing a solution to Eq.~(\ref{ms}). Using this solution, we derive
 $\mathcal{P}_\zeta[k_i; \vec{\theta}]$ as in Eq.~(\ref{ps}) for a specific range of $k$, and employ parallel computations. Next, we identify the set of parameters $\vec{\theta}$ in the chosen model  whose power spectrum matches $\mathcal{P}^{\text{rec}}_\zeta(k)$.

In the case of the E-model under consideration, the procedure described above allowed us to get the parameter Set 3 in Table 1 (see Sec.~7), starting from $\Omega_{\text{BPL}}$ shown in Fig.~\ref{fig6}.

Knowing the inflationary model underlying a given GW signal, allows us to verify whether this model is capable of reproducing the observed signal without conflicting with CMB measurements or quantum corrections. Such verification requires knowledge of the power spectrum beyond the scales directly reconstructed from the GW signal.

\section{The results}

In this Section, we present our main findings.

Three sets of fine-tuned parameters in our E-model, required for efficient PBHs production relevant to DM and GW, are collected in Table 1.

\begin{table}[H]\label{pars}
\centering
\begin{tabular}{|c|c|c|c|}
\hline
\textbf{E-Model} & \textbf{Set 1} & \textbf{Set 2} & \textbf{Set 3} \\ \hline
$\alpha$         & $0.7425$         & $0.74$          & $0.77218$          \\ \hline
$\phi_i$         & $-\,0.611325$      & $-\,0.6112$       & $-\,0.6415$        \\ \hline
$\sigma$         & $0.012137$      & $0.0131$        & $0.0125$         \\ \hline
$\theta$         & $-\,7.29\cdot10^{-7}$   & $-\,9.17\cdot10^{-7}$  & $-\,7.37\cdot10^{-7}$ \\ \hline
$(M/M_{\rm Pl})^2$              & $6.3\cdot10^{-10}$ & $7.2\cdot10^{-10}$ & $6.3\cdot10^{-10}$ \\ \hline
\end{tabular}
\caption{The fine-tuned parameters for the E-model of inflation and PBH DM.}
\end{table}
\vspace{-5mm}

With those parameters, the E-model satisfies the CMB constraints and allows a production of PBHs in the asteroid-mass (atomic size) window, where those PBHs may  account for a significant part (or the whole) of DM.\footnote{The PBH abundances vary, being dependent on the approach used (see Ref.~\cite{Pi:2024ert} for a review).} Gravitational collapse of scalar perturbations leading to  the PBH formation induces the stochastic GW that may be detectable  by the future space-based gravitational interferometers such as LISA \cite{LISA:2017pwj,LISACosmologyWorkingGroup:2024hsc,LISACosmologyWorkingGroup:2025vdz}, TAIJI \cite{Gong:2014mca}, TianQin \cite{TianQin:2015yph} and DECIGO \cite{Kudoh:2005as}, see 
Refs.~\cite{Bartolo:2018evs,Garcia-Bellido:2017aan,Cai:2018dig} for more about them.

We used the linear approximation for computing the CMB tilt $n_s$ of scalar perturbations and the CMB tensor-to-scalar ratio $r$ from the Hubble-flow parameters at the pivot CMB scale $k_*=0.05~{\rm Mpc}^{-1}$ at the horizon crossing,
\begin{equation}
n_s=1-2 \epsilon-\eta~, \quad r=16 \epsilon~.
\end{equation}
In the E-Model, it leads to approximately 60 e-folds. The current measurements of the CMB give \cite{Planck:2018jri}:\footnote{The running of the scalar spectral index $n_s$, usually denoted by $\alpha_s$, is also an observable. Its observational bounds are satisfied for each set of the parameters in Table 1.}
\begin{equation}
n_s=0.9649 \pm 0.0042 \quad(68 \% \text { C.L.})~, \quad {\rm and}\quad r<0.032 \quad(95 \% \text { C.L.)}~.
\end{equation}

The PBH masses can be estimated as \cite{Inomata:2017okj,Inomata:2018cht}
\begin{equation}
M_{\mathrm{PBH}}(k) \simeq 10^{20}\left(\frac{7 \cdot 10^{12}}{k \cdot  \mathrm{Mpc}}\right)^2 \mathrm{g}~.
\end{equation}

The key observables and the relative values of the one-loop correction (1LC) are given in Table 2.

\begin{table}[H]
\centering
\begin{tabular}{|c|c|c|c|c|c|c|c|}
\hline
\textbf{E-Model} & $n_s$ & $r$  & $h$ &$\Delta N_{\rm USR}$& $M_{\rm PBH},\mathrm{~g}$ & $\mathcal{P}_\zeta(k_{\rm peak})$ & $\delta_{\rm 1L}$, \% \\ \hline
\textbf{  Set 1} & $0.9649$ & $0.01466$ & $-\,1.472$ & $2.674$ & $2.2\cdot 10^{21}$ & $10^{-3}$ & $1.31$ \\ \hline
\textbf{  Set 2} & $0.9649$ & $0.01685$ & $-\,1.435$ & $2.838$ & $7.1\cdot 10^{22}$ & $0.5 \cdot 10^{-2}$ & $2.96$ \\ \hline
\textbf{  Set 3} & $0.9649$ & $0.014323$ & $-\,1.471$ & $2.687$ & $4.6\cdot 10^{18}$ & $ 10^{-3}$ & $1.33$ \\ \hline
\end{tabular}
\caption{The CMB tilts $n_s$ and $r$, the sharpness parameter $h$, the USR phase duration $\Delta N_{\rm USR}$, the PBH masses $M_{\rm PBH}$, the peak amplitude $\mathcal{P}_\zeta(k_{\rm peak})$, and the relative (against the tree level contribution) one-loop correction $\delta_{\rm 1L}$ for the parameter sets in Table 1.}
\end{table}

We also evaluated the $\mu$-type distortion \cite{Zeldovich:1969ff,Chluba:2011hw} of the CMB in our model, which places the additional constraints on the primordial power spectrum at smaller scales up to~$k\sim 10^4\,\mathrm{Mpc}^{-1}$. The $\mu$-distortion can be estimated by the integral \cite{Chluba:2015bqa, Schoneberg:2020nyg, Unal:2020mts, Pajer:2012vz, Tagliazucchi:2023dai}
\begin{equation}
	\mu \approx \int\limits_{k_{\min }}^{\infty} \frac{\mathrm{d} k}{k}\,\mathcal{P}_\zeta(k)\,W_\mu(k)~,
\end{equation}
where the analytic window function $W_\mu(k)$ is given by \cite{Pajer:2012vz}
\begin{equation}
	W_\mu(k) \approx 2.27\left\{\exp \left[-\left(\frac{k}{1360}\right
	)^2 \left(1+\left(\frac{k}{260}\right)^{0.3}+\frac{k}{340}\right)^{-1} \, \right]-\exp \left[-\left(\frac{k}{32}\right)^2\right]\right\}\,,
\end{equation}
and $k_{\min}\simeq 1 \,\mathrm{Mpc}^{-1}$. The current upper bound from COBE/FIRAS observations \cite{Fixsen:1996nj} is given by 
$\mu < 9 \cdot 10^{-5}$ ($95\%\, \text{C.L.}$). In our model, for each parameter set in Table 1, we find $\mu \simeq 10^{-10}$ that is significantly below the observational limit.

The results of our numerical calculations, related to the data in Table 1 and Table 2, are given in Figs. 1--6. In particular, we found the one-loop correction (1LC) in our E-model does not exceed a few percent, being primarily depending upon the amplitude of the peak in the scalar power spectrum. These results are consistent with the conclusions in Ref. \cite{Davies:2023hhn}.

We also found that extending the limits of integration in Eq.~(\ref{1LC}) has a negligible effect on the results, thus confirming the dominant contribution to the 1LC comes from the cubic interaction during the period when $\eta^{\prime}$ is large.  An impact of the quartic interaction deserves future research.

The inverse reconstruction of the parameters in the E-model from the GW spectrum  demonstrates its potential as a reasonable framework for interpreting future GW measurements.

\section{Conclusion} 

We conclude with a few relevant comments.

The E-model defined by Eq.~(\ref{newpar}) is a deformation of the Starobinsky model of inflation~\cite{Starobinsky:1980te}.
Like any other viable single-large-field model of inflation, it is tightly constrained but is not ruled out by CMB measurements of
$(A_s,n_s,r)$, and the Swampland conjectures \cite{Scalisi:2018eaz}. The E-type $\a$-attractor model investigated in this paper is different from the T-type $\a$-attractor models, and from other single-field models of inflation and PBH production studied in the literature 
\cite{Martin:2013tda,Karam:2022nym,Martin:2024qnn}. 

The 1LC to the power spectrum of  our E-model was found to be merely a few percent against the tree-level
contribution, which further supports the validity of the model. The estimates of the two-loop contribution (2LC) in the literature \cite{Saburov:2024und,Firouzjahi:2024sce}  suggest the 2LC to be roughly the 1LC squared, which is negligible in our E-model also.
An impact of the quartic interactions to the 1LC in our model deserves further studies, while those interactions may further 
reduce the 1LC contribution, see Refs.~\cite{Fumagalli:2023hpa,Braglia:2024zsl,Kawaguchi:2024rsv,Fumagalli:2024jzz}.

Reheating (particle production after inflation) in our E-model is the same as that in the Starobinsky model, which was well studied in the literature, see e.g., Ref.~\cite{Jeong:2023zrv} and references therein, so we do not expect tensions with standard cosmology
\cite{Allegrini:2024ooy}.

Simple single-field SR models of inflation were motivated by the almost scale-invariant CMB spectrum and approximately Gaussian fluctuations. Though being tightly constrained, those models have high predictive power, and they may be either confirmed or falsified by
future cosmological measurements. Non-Gaussianities (or non-linear effects) at peak scales are expected to be of the local type with the small amplitude of $f_{\rm NL} \sim{\cal O}(0.1)$ \cite{Iacconi:2024hmg}, even though this key estimate may be not enough to claim validity of the perturbative approach. Non-perturbative non-Gaussianities \cite{Vennin:2024yzl,Caravano:2024moy} are expected due to stochastic effects during inflation but their inclusion is beyond the scope of this paper.

We avoided presenting our results about a specific PBH fraction in DM, based on our E-model, because the PBH abundance is highly sensitive to the peak in the power spectrum, while the standard (Press-Schechter) formalism for its evaluation does not produce reliable results.

\section*{Acknowledgements}

The authors are grateful to Matteo Braglia, Angelo Caravano, Mattia Cielo, Gabriele Franciolini, Jacopo Fumagalli, Antonio Junior Iovino,
Kazunori Kohri, Haidar Sheikhahmadi  and Shinji Tsujikawa for correspondence.

DF and SVK were partially supported by Tomsk State University under the development program Priority-2030. DF was supported by the Foundation for Advancement of Theoretical Physics and Mathematics "BASIS".  SVK was also supported by Tokyo Metropolitan University, the Japanese Society for Promotion of Science under the grant No.~22K03624, and the World Premier International Research Center Initiative, MEXT, Japan.

\bibliographystyle{utphys}
\bibliography{Bibliography.bib}

\end{document}